\documentclass[aps,pra,twocolumn,unsortedaddress]{revtex4-1}

\usepackage{graphicx}
 \usepackage{amsmath}
 \usepackage{multirow}

\begin{document}

\title{Antihydrogen level population evolution: impact of positron plasma length}

\author{B. Radics}
\email{balint.radics@riken.jp}
\author{Y. Yamazaki}
\affiliation{Atomic Physics Laboratory, RIKEN, Saitama 351-0198, Japan}

\begin{abstract}

Antihydrogen is produced by mixing an antiproton and a positron plasma in a cryogenic electromagnetic trap. The dominant antihydrogen formation mechanism is three-body recombination, while the subsequent level population 
evolution is governed by various processes, mainly collisional (de)excitation, ionisation and radiative decay. In this work the impact of various
positron plasma lengths on the level population evolution is investigated. The main interest is the ground-state antihydrogen atom yield. It is found that the ground state level population shows different power-law behaviour at short or longer positron plasma lengths.
\end{abstract}


\maketitle

%----------------------------------------------------------------------------------------
%	ARTICLE CONTENTS
%----------------------------------------------------------------------------------------

\section{Introduction}

Today antihydrogen atoms can be formed, trapped and observed routinely by various experiments \cite{WOelert} \cite{Amoretti_Nat2002} \cite{Gabrielse_PRL2002} \cite{Andresen_Nat2010} \cite{Alpha_NatPhys2011} \cite{Gabrielse_PRL2012}. Very recently antihydrogen beams prepared for in-flight spectroscopy has been reported \cite{Kuroda_NatComm2014}. The primary motivation of these experiments
is to compare the spectroscopic properties of the antihydrogen atom to those of the hydrogen atom, which is a direct test of the CPT symmetry. \\
The standard method of formation of antihydrogen is to produce a mixture of antiproton and positron plasmas in an electromagnetic trap
under cryogenic conditions. Experimentally the important spectroscopic measurements are to be performed on ground-state antihydrogen
atoms, while the antihydrogen atoms are created dominantly via the three-body recombination process ($e^{+} + e^{+} + \bar{p} \rightarrow \bar{H} + e^{+}$), which primarily populates highly excited Rydberg states. In order to reach ground-state experiments either trap the antiatoms or prepare them for in-flight spectroscopy. In both cases the radiative decay process eventually relaxes a fraction of the level population towards ground-state. However, the initial three-body recombination process, which takes place inside the positron plasma, populates primarily highly excited levels, therefore the rate of populating the ground-state level is also impacted by the scattering processes inside the plasma.
\\
A key element for efficient antihydrogen production is to control the positron plasma conditions and to repopulate high Rydberg states to intermediate and low excited states. The rate of the three-body recombination process is proportional to the positron 
density to the second power, while the proportionality for the rates of other processes is typically to the zeroth (radiative deexcitation)
or the first power (collisional (de)excitation) of the positron density (see \cite{Robicheaux_2004} and \cite{Pohl_2008} and references therein). The rates of the various processes in magnetic field has been studied in great lengths previously \cite{Glinsky_1991} \cite{Robicheaux_PRA2006} \cite{Topcu_FR_2006} \cite{Hurt_JPB2008}. A review has been published by Robicheaux \cite{Robicheaux_review}. Bass and Dubin showed calculation results for level population evolution under cryogenic and highly magnetised conditions \cite{BassDubin_2009}. They characterised this regime by a magnetisation parameter, $\chi \equiv \bar{v}/b\Omega_{c} = 0.0018(T_e/4\; K)^{3/2}/(B/6\;T)$, where $T_e$ is positron temperature, $B$ magnetic field strength, $\bar{v}$ positron thermal speed, $b$ classical distance of closest approach and $\Omega_{c}$ positron cyclotron frequency.  In their work the magnetisation parameter range explored was from $\chi = 0$ (infinite magnetic field) to $\chi = 0.005$ (roughly a combination of $T_e = 10$ K positron temperature and $B = 10$ T magnetic field strength). The impact of positron plasma conditions on the ground-state antihydrogen yield under a broad range of positron plasma density and temperature scales was published by Radics, Murtagh, Yamazaki and Robicheaux \cite{Radics_PRA2014} covering the magnetisation parameter range of $0.04 \leq \chi \leq 3.5$, which corresponds to recent experimentally achievable conditions in current antihydrogen experiments ($20$ K $< T_e < 300$ K and $B \simeq 2-3$ T).\\
Because of the crucial role of the positron plasma conditions in antihydrogen formation, in this work we extend our results from \cite{Radics_PRA2014} and investigate the impact of various positron plasma lengths on the level population evolution and on the useful ground-state antihydrogen yield, in the same magnetisation range, $0.04 \leq \chi \leq 3.5$.\\
This paper is organised as follows. In section~\ref{sec:LevPopMod} we briefly summarise the antihydrogen level population evolution model. Then in section~\ref{sec:Discussion} we present our findings on the level population evolution when scanning with various plasma lengths. Finally the results are concluded in 
section~\ref{sec:Conclusion}.

%------------------------------------------------

\section{\label{sec:LevPopMod}Level population model}

The antihydrogen level population evolution model describes the passage of a number of antiprotons through a cloud of positrons. During their passage the antiprotons may recombine with positrons, forming antihydrogen atoms, and afterwards they may take part in various other scattering and decay processes. The evolution is based on a set of coupled differential equations evolving the system in time. The time evolution of the population of each quantum state, $N(i)$, is governed
by the rate equations
\begin{equation}
\label{eq:difflevpop}
\begin{split}
\frac{dN(i)}{dt} &= [C_{rr}(i) + C_{tbr}(i)n_{e}]n_{e}N_{p} - C_{ion}(i)n_{e}N(i) \\
& + \sum_{j \neq i}[C_{col}(j, i)n_{e} + C_{str}(j, i)]N(j) \\
& - N(i)\sum_{j\neq i}[C_{col}(i,j)n_{e} + C_{str}(i,j)],
\end{split}
\end{equation}
while the number of bare antiprotons fulfills the rate equation
\begin{equation}
\frac{dN_{p}}{dt} = \sum_{i} \left( C_{ion}(i)n_{e}N(i) - [C_{rr}(i) + C_{tbr}(i)n_{e}]n_{e}N_{p} \right),
\end{equation}
where $N_{p}$ is the number of antiprotons, $n_e$ is the density of positrons, $C_{rr}(i)$ and $C_{tbr}(i)$ denote rate coefficients for radiative and three-body recombination to state $i$ respectively, $C_{ion}(i)$ denotes ionisation by positron impact from bound state $i$, $C_{col}(i,j)$ denotes collisional excitation or deexcitation by positron impact from state $i$ to state $j$ and $C_{str}(i,j)$ denotes spontaneous or stimulated transitions due to presence of a radiation field. The thermal equilibrium rate coefficients has been calculated by classical-trajectory Monte Carlo method, as described in \cite{Radics_PRA2014}. The set of differential equations are solved using the Bulirsch-Stoer method \cite{GSL}.\\
Initially, the model starts with empty level population for each quantum state and a given number of antiprotons. Then during the subsequent time integration the population of the quantum states of the antihydrogen atoms is being filled and distributed by recombination, scattering and decay processes, the rates of which are all contained in the coefficients of the various terms in the coupled differential equations.

%------------------------------------------------

\section{\label{sec:Discussion}Discussion}

\subsection{Impact of positron plasma length}

\begin{figure}[ht]
\vspace{20pt}
\includegraphics[width=8.0cm]{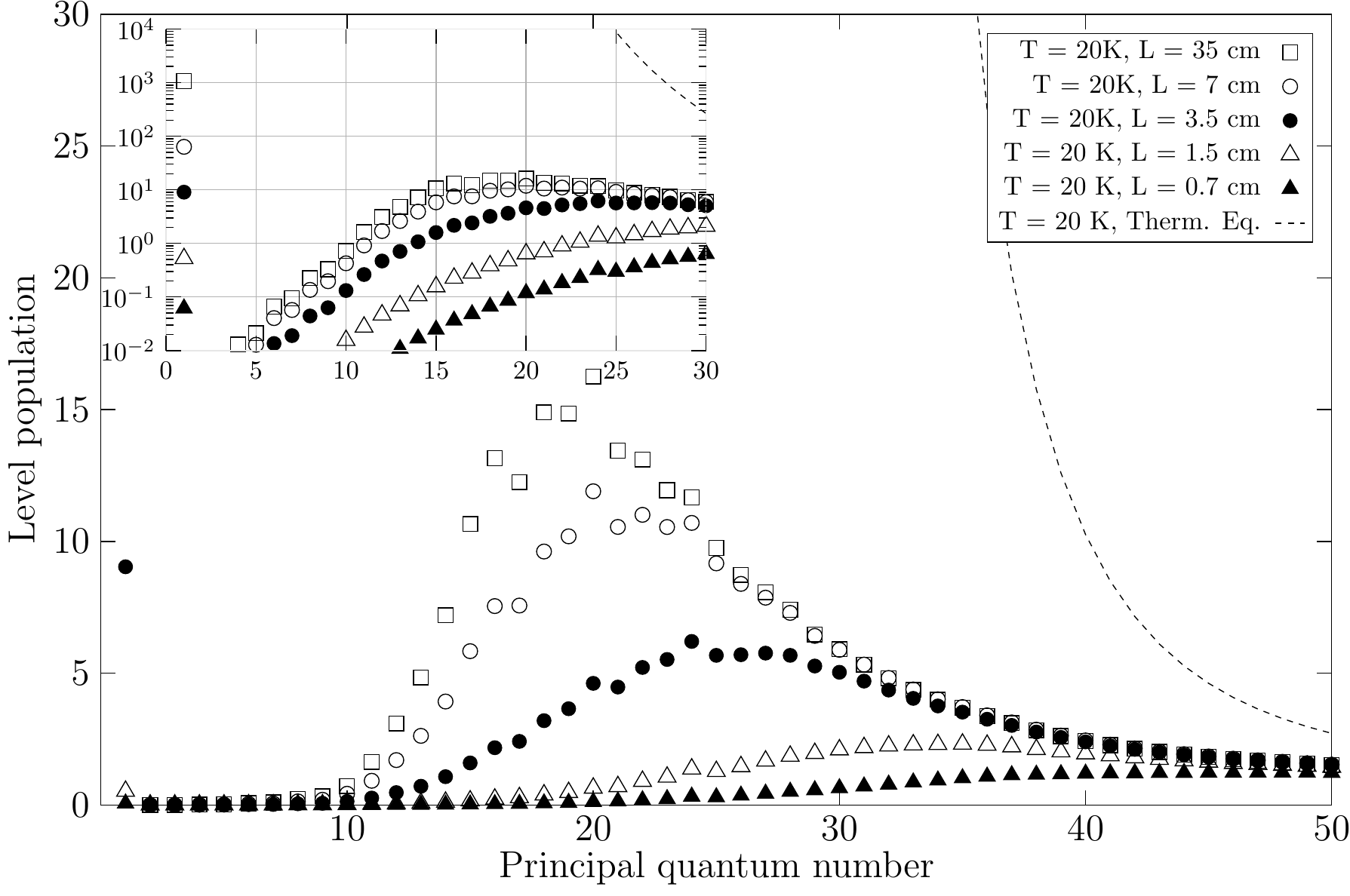}
\caption{Antihydrogen level population distribution for various plasma lengths, using temperature $T= 20$ K, density $n_e=10^{14}$m$^{-3}$, magnetic field strength $B = 2$ T. The subplot shows the same distribution on logarithmic scale, and zoomed onto the principal quantum number range $n = 1-30$.}
\label{LevPop_T20K}
\end{figure}

\begin{figure}[ht]
\vspace{20pt}
\includegraphics[width=8.0cm]{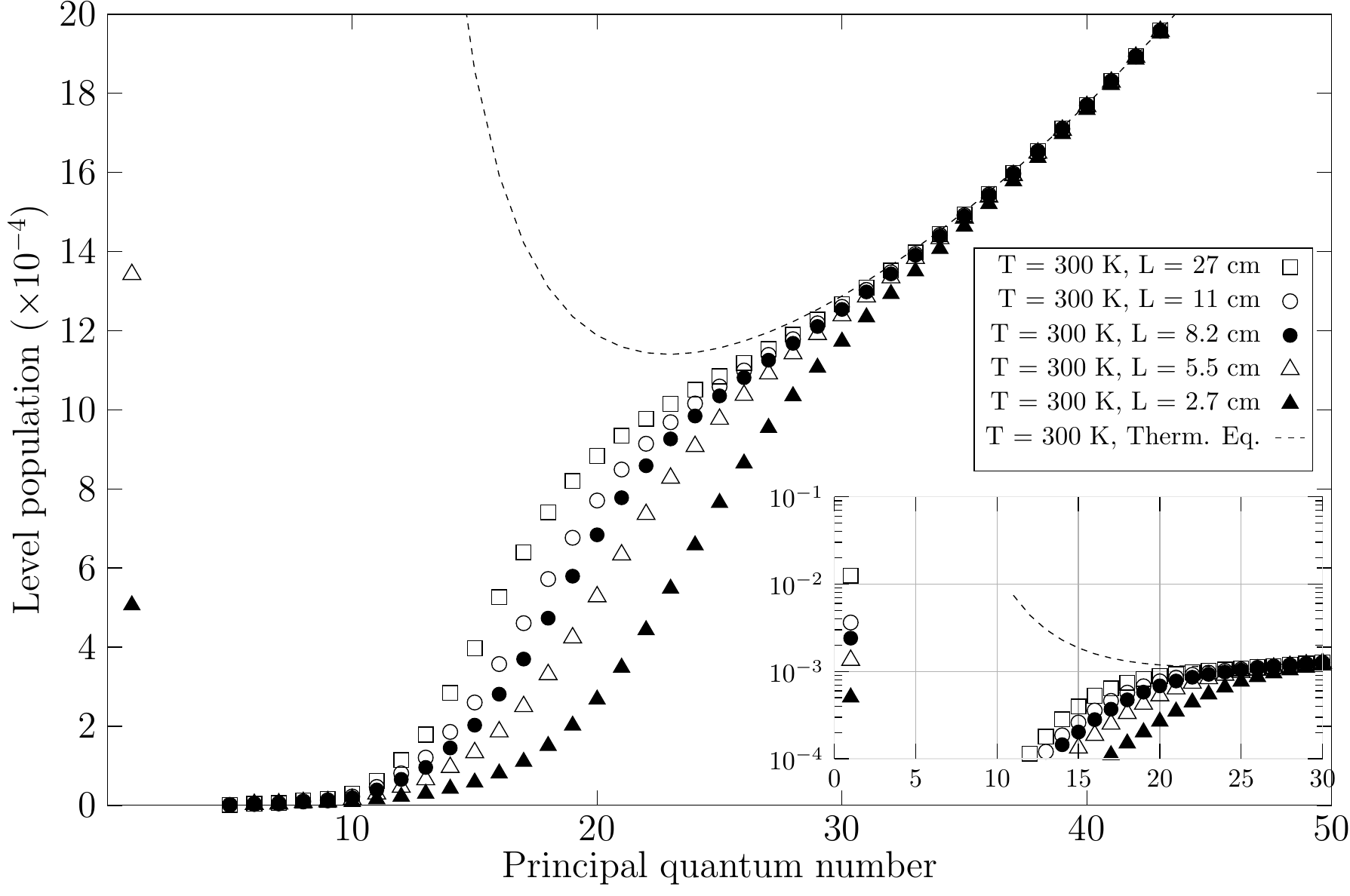}
\caption{Antihydrogen level population distribution for various plasma lengths, using temperature $T = 300$ K, positron density $n_e=10^{14}$m$^{-3}$, magnetic field strength $B = 2$ T. The subplot shows the same distribution on logarithmic scale, and zoomed onto the principal quantum number range $n = 1-30$.}
\label{LevPop_T300K}
\end{figure}

\begin{figure}[ht]
\vspace{20pt}
\includegraphics[width=8.0cm]{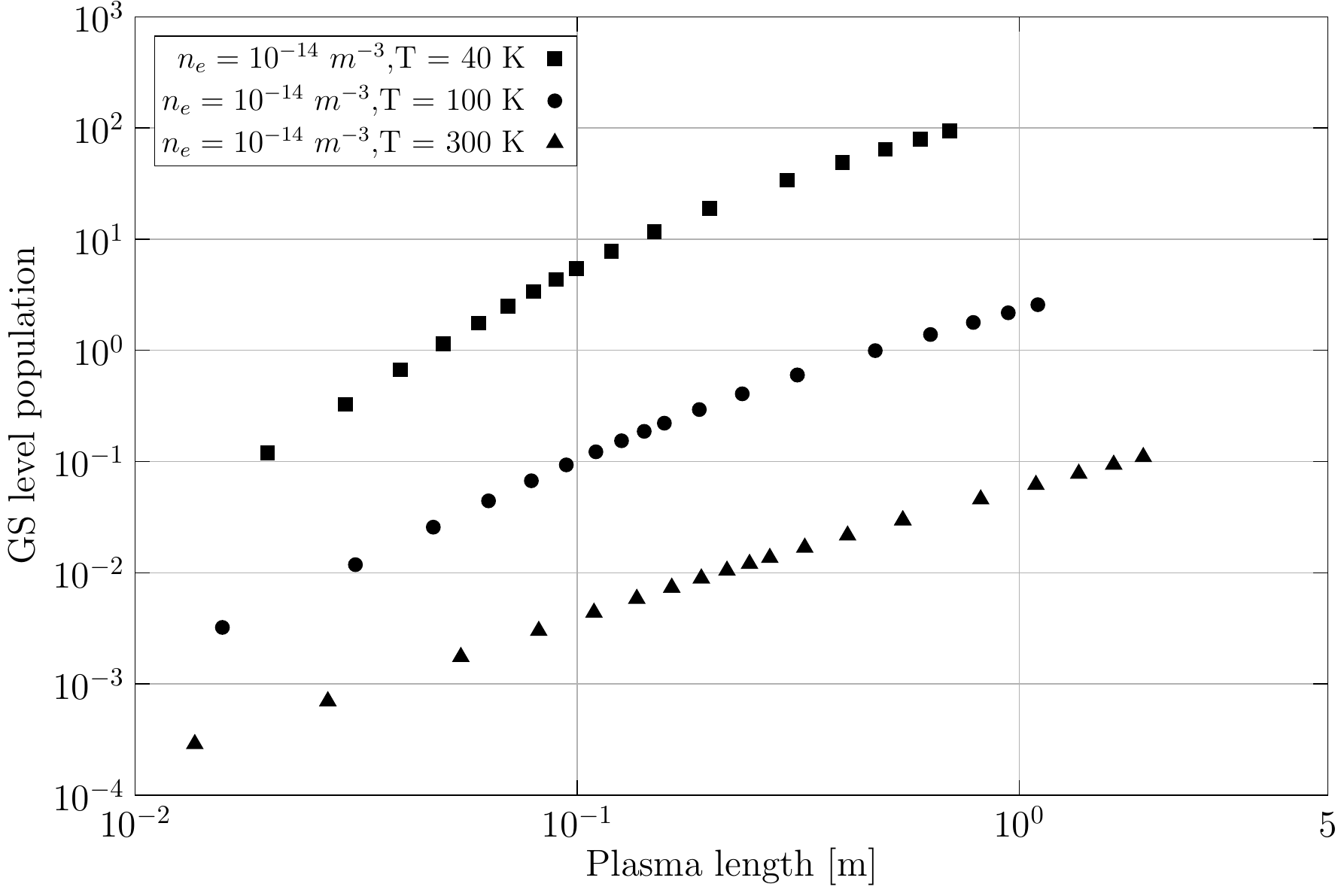}
\caption{Ground-state antihydrogen level population ($n \leq 15$) as a function of plasma length, using various temperature values at a fixed positron density $n_e=10^{14}$m$^{-3}$. }
\label{LengthScan}
\end{figure}

\begin{figure}[ht]
\vspace{20pt}
\includegraphics[width=8.0cm]{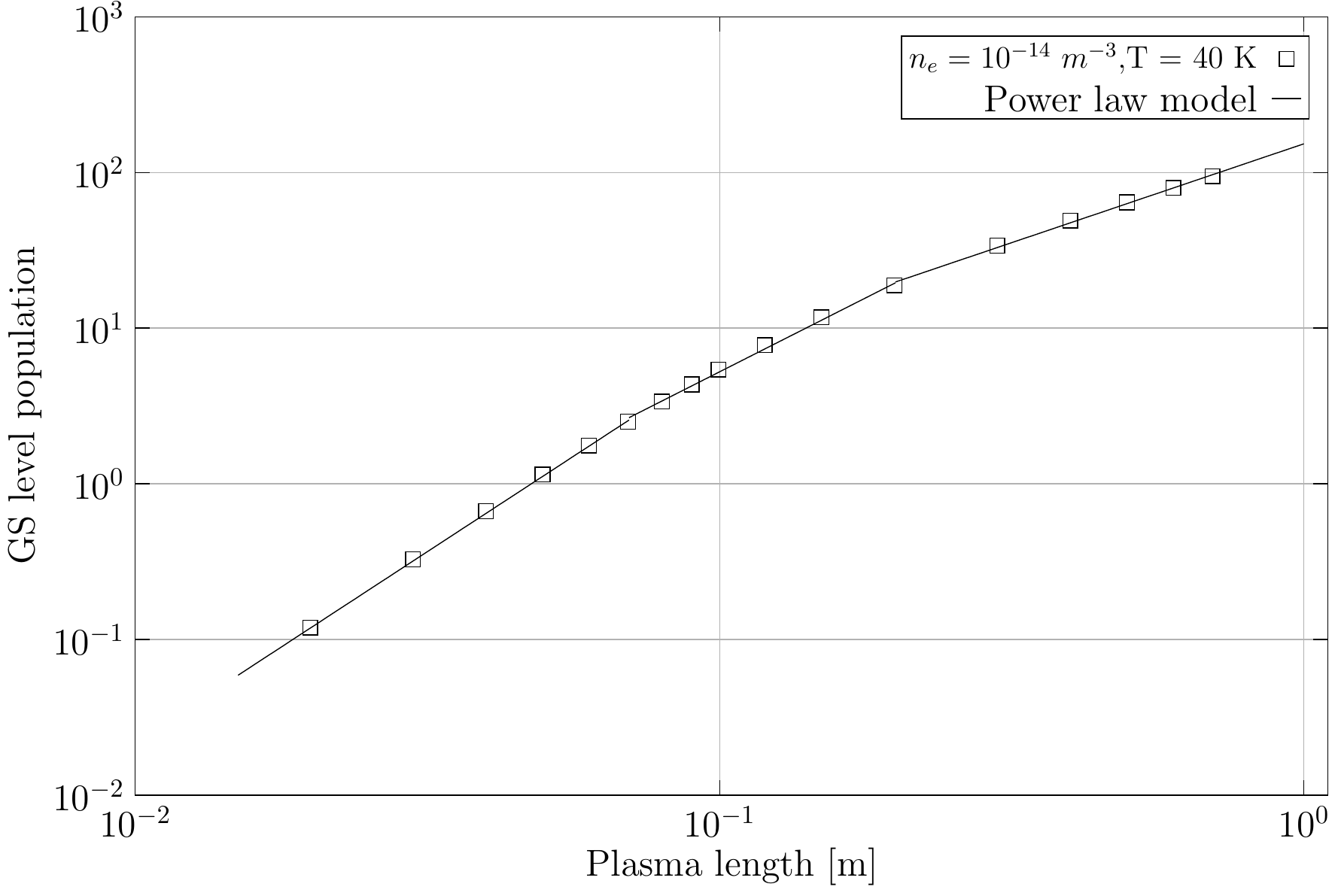}
\caption{Power-law fit using the parametrisation in Equation~\ref{eq:gs_vs_plength} for plasma temperature $T = 40$ K and at fixed positron density $n_e = 10^{14}$ $\mathrm{m}^{-3}$ and magnetic field strength $B = 2$ T. The fit values are presented in Table~\ref{fit_tab} .}
\label{fit_T40K}
\end{figure}

The effect of different positron plasma lengths can be factorised into the evolution time, which is the time the antiprotons and antihydrogen atoms spend inside the positron plasma. This time depends on the velocity of antiprotons and the length of the plasma. We assume in this model that the temperature of the antiproton and positron plasma is equalised quickly once they start to mix. We note that our model could in principle be easily adapted to a non-equilibrium thermal evolution code by assuming local thermal equilibrium, either in space of time, re-evaluating the level population after each step, and subsequently feeding the level population obtained from a previous step to the next step during the calculation. Because of the assumption of the thermal equilibrium the evolution time and subsequently the plasma lengths traversed by the antiprotons are slightly different at low or high temperatures. Typical time scale ranges for a single passage of antiprotons through the positron clouds with various lengths are in the order of $\simeq 1-100$ $\mu$s. In the simulation positron temperatures in the range $T_e = 20 - 300$ K were used with a fixed positron density of $n_e=10^{14}$m$^{-3}$, a $B = 2$ T magnetic field strength, and with a total number of antiprotons of $N_p = 10^{5}$. Because of the assumption of thermal equlibrium the temperature values we use interchangeably for positrons or antiprotons. The level population distributions obtained for example at $T_e = 20$ K and at $T_e = 300$ K are shown in Figure~\ref{LevPop_T20K} and Figure~\ref{LevPop_T300K}, respectively, with subfigures showing the level population on logarithmic scale. There are several interesting features to note from these distributions. The longer the plasma length the more the $n = 1$ ground-state antihydrogen levels are populated, which is understood since more time is allowed for the population to evolve and for low level quantum states to decay to ground-state. At $T_e = 20$ K temperature the ground-state population is found to increase around four orders of magnitude in the range of plasma lengths from $L = 0.7$ cm to $L = 35$ cm. While at $T_e = 300$ K the ground-state population increases only around two orders of magnitude in a similar plasma length range. It is also observed that at $n\simeq20$ principal quantum number the states become more populated with increasing plasma length. This effect indicates that the quantum state distribution at high-n states and at very long plasma lengths starts to evolve, although very slowly, towards the thermal equilibrium level population distribution, given by the Saha-Boltzmann relation, $N_{th}(i) = N_{p} n_{e} n_{i}^{2} \Lambda^{3} e^{\frac{E(i)}{k_{B}T_{e}}}$, while states below $n \simeq 20$ principal quantum number decay radiatively to ground-state quickly. In the Saha-Boltzmann formula $N_p$ is the number of antiprotons used in the simulation, $n_{i}$ is the principal quantum number of state $i$, $\Lambda=h/\sqrt{2\pi m_e k_{B} T_{e}}$ is the thermal de Broglie wavelength of the positron, $m_{e}$ is the positron mass, $E(i)$ is the binding energy of state $i$, $h$ is the Planck's constant and $k_B$ is the Boltzmann constant. The thermal equilibrium level population distribution is indicated with dashed lines on Figure~\ref{LevPop_T20K} and Figure~\ref{LevPop_T300K} using the fixed positron density value $n_e=10^{14}$m$^{-3}$. It is also noted that the peak position of the build-up of the level population shifts towards lower principal quantum numbers with increasing plasma lengths, subsequently populating more quantum states closer to the ground-state. \\
In this paper our aim is to study the experimentally useful ground-state level population, therefore we adopt the number of states in $n \leq 15$ quantum state as a measure of useful states as they decay to ground-state within 1 ms. Using this measure the number of antihydrogen atoms with states $n \leq 15$ are presented as a function of the plasma length on Figure~\ref{LengthScan}.
With a composite power-law model we could parametrise the obtained ground-state level population dependence on the plasma length for various temperature values,
\begin{equation}
\label{eq:gs_vs_plength}
   f(x; ...)= 
\begin{cases}
    A x^{a},& \text{if } x = [l^{A}_{\mathrm{min}}, l^{A}_{\mathrm{max}}]\\
    B x^{b},& \text{if } x = [l^{B}_{\mathrm{min}}, l^{B}_{\mathrm{max}}] ,\\
    C x^{c},& \text{if } x = [l^{C}_{\mathrm{min}}, l^{C}_{\mathrm{max}}]\\
    \end{cases}
\end{equation}
where the normalisation constants and power-law fit parameters are $A$, $B$, $C$ and $a$, $b$, $c$, respectively, and the plasma length ranges for each power-law behaviour is labelled as $[l_{\mathrm{min}}, l_{\mathrm{max}}]$. We found that a different power-law behaviour occurs at shorter or longer length scales ranging from order of a centimeter to the order of a meter. We were only able to fit the ground-state level population dependence using three rough length ranges, which we label with $A$, $B$ and $C$, each of them showing a different power-law behaviour. These length ranges are $[l^{A}_{min}, l^{A}_{max}] = [0.015, 0.07]$, $[l^{B}_{min}, l^{B}_{max}] = [0.07, 0.2]$ and $[l^{C}_{min}, l^{C}_{max}] = [0.2, 1.0]$ in units of meter. The parameter values obtained during the fit for the fixed positron density value $n_e = 10^{14}$ $\mathrm{m}^{-3}$ are shown in Table~\ref{fit_tab}. An example fit result at $T = 40$ K is shown in Figure~\ref{fit_T40K}.  \\
\begin{table}[ht]
	\begin{tabular}{| c | c | c | c | c | c | c | } \hline
	\textbf{Temperature [K]} & $\mathbf{A}$ & $\mathbf{a}$ & $\mathbf{B}$ & $\mathbf{b}$ & $\mathbf{C}$ & $\mathbf{c}$ \\ \hline \hline
	40 &  1734 & 2.45 & 416 & 1.90 &  153 & 1.27 \\ \hline
	100 &  8.48 & 1.90 & 5.09 & 1.70 &  2.38 & 1.21 \\ \hline
	300 &  0.09 & 1.35 & 0.07 & 1.23   & 0.06 &  1.08 \\ \hline 
	\end{tabular}  
	\caption{Fit parameter values for composite power-law Equation~\ref{eq:gs_vs_plength} for temperature values $T=40$ K, $100$ K and $300$ K, and fixed positron density value $n_e = 10^{14}$ $\mathrm{m}^{-3}$, and for
	positron plasma length scales $[l^{A}_{min}, l^{A}_{max}] = [0.015, 0.07]$, $[l^{B}_{min}, l^{B}_{max}] = [0.07, 0.2]$, $[l^{C}_{min}, l^{C}_{max}] = [0.2, 1.0]$ in meters .}
	\label{fit_tab}
\end{table}
%	  \bf{Plasma parameter} & $l^{A}_{\mathrm{min}} - l^{A}_{\mathrm{max}}$ m & A & a & B & b & C & c \\ \hline 
The power values from the results shown in Table~\ref{fit_tab} suggest that towards longer plasma lengths the rate of filling the antihydrogen ground-state level population does not follow a constant power dependence but rather the power gets smaller with longer plasma lengths. This seems to be consistent with the build-up of population at principal quantum number $n \simeq 20$ as discussed previously. The decrease in the power with plasma length is also found at higher positron plasma temperature, although it is less pronounced. While at $T = 40$ K the power is found to decrease from $a = 2.45$ to as low as $c = 1.27$, at $T = 300$ K it decreases from $a = 1.35$ to only $c = 1.08$ with longer plasma lengths. Such a decrease in power suggests that longer time is available for collisional processes to take place, the rate of which scales with lower power on positron plasma density. However, over two orders of magnitude increase in plasma length the ground-state level population is found to increase by two-three orders of magnitude, independent of the plasma temperature, which may enhance the precision of experiments pursuing spectroscopic measurements using ground-state antihydrogen atoms. Although such an extremely long positron plasma may be challenging to be realized and to be kept stable experimentally.

%------------------------------------------------

\section{\label{sec:Conclusion}Conclusion}

In this work the antihydrogen level population dependence was investigated as a function of the positron plasma length in the plasma magnetisation parameter range of $0.04 \leq \chi \leq 3.5$, which corresponds to recent experimentally achievable conditions in current antihydrogen experiments. It was found that a composite power-law parametrisation can fit the rate of ground-state level population, using different power-law values in various plasma length scales. The power-law behaviour was found to decrease with increasing plasma lengths at each of the investigated temperature scales in the range $T_e = 40-300$ K. The results also show that a two orders of magnitude longer plasma could increase the ground-state level population of antihydrogen by two-three orders of magnitude, independent of the plasma temperature.

%------------------------------------------------
\begin{acknowledgments}

This work was supported by the Grant-in-Aid for Specially Promoted Research (no. 24000008) of the Japanese Ministry of Education, Culture, Sports, Science and Technology (Monbukakagu-sho), Special Research Projects for Basic Science of RIKEN, RIKEN programme for young scientists. 

\end{acknowledgments}

\newpage
% Create the reference section using BibTeX:
\bibliography{Bibliography}

\end{document}